Body of the paper

\documentstyle[12pt]{article}
\begin{document}
\baselineskip=24pt
\def\s{\section}
\def\ss{\subsection}
\def\sss{\subsubsection}
\def\ni{\noindent}
\def\hf{\hfill\break}
\def\i{\item}
\def\mclc{Mol. Cryst. Liq. Cryst.}
\def\mclcl{Mol. Cryst. Liq. Cryst. Lett.}
\def\jp{J. de Phys.}
\def\jpl{J. de Phys. Lett.}
\def\lc{Liq. Cryst.}
\def\pr{Phys. Rev.}
\def\prl{Phys. Rev. Lett.}
\def\be{begin{equation}}
\def\ee{end{equation}}
\def\ni{\noindent}

\title{Observation of the Smectic C -- Smectic I Critical Point}
\author{S. Krishna Prasad and D.S. Shankar Rao\\
Raman Research Institute, Bangalore 560080, INDIA\\
S.Chandrasekhar\\
Centre for Liquid Crystal Research, P.O.Box 1329, Jalahalli, Bangalore 560013,
INDIA\\
M.E.Neubert\\
Liquid Crystal Institute, Kent State University, Kent, Ohio 44240, U.S.A\\
J.W.Goodby\\
University of Hull, Hull HU6 7RX U.K.}
\date{}
\maketitle
\begin{abstract}
We report the  first observation of the smectic C--smectic I
(C--I) critical point by Xray diffraction studies on a binary
system.  This is in confirmity with the theoretical idea of
Nelson and Halperin that coupling to the molecular tilt should
induce hexatic order even in the C phase and as such both C and
I (a tilted hexatic phase) should have the same symmetry. The
results provide evidence in support of the recent theory of
Defontaines and Prost proposing a new universality class for
critical points in layered systems.
\end{abstract}

PACS numbers 61.30, 64.60.Fr, 64.70.M \\~\\
\ni
Smectic C (C) and Smectic I (I) are liquid crystalline phases
in which the molecules are tilted with respect to the layer
normal.  In the I phase the tilt order is supplemented by a
six-fold bond orientational (BO) order. Nelson and
Halperin\cite{ci1} predicted that coupling to molecular tilt should
induce long-range BO order even in the C phase. This has been
confirmed by synchrotron Xray studies\cite{ci2}. Since only the magnitude
of BO order is different, both C and I phases have the same point
group symmetry. An immediate consequence of this feature is that
there can be either a first-order transition or no transition
but a continuous evolution from one phase to the other. The existence
of first-order transition for some systems
\cite{ci3}-\cite{ci5}, together with the continuous evolution
situation for others \cite{ci2,ci5}, implies that there should
exist a critical point (CP) in the phase diagram. However, the
occurrence of such a point has not been experimentally observed
so far. These studies  have been made more significant by the
recent predictions of Defontaines and Prost (DP)\cite{ci6}.  The
DP theory, based on the concept of the existence of
qualitatively different spatial directions in layered systems,
expects the CP in smectic systems to belong to the same but new
universality class.

Two of the critical points considered to be in the DP
universality class have already been observed
experimentally.  They are the Smectic A$_d$--Smectic A$_2$ CP\cite{ci7} and
the electric field induced Smectic C$^*$--Smectic C CP\cite{ci8}. The
critical exponents obtained in these cases\cite{ci8,ci9} suggest that
the universal description of CP in layered systems is a definite
possibility.

In this letter we present the results of Xray studies on binary
mixtures of Terepthal-bis-decylaniline\cite{ci9b} (TB10A for short)
and 4-n-decyloxy biphenyl 4-(2$^\prime$-methylbutyl)
benzoate\cite{ci9c} (C10 for short) which have led to the first
observation of the C--I critical point. We demonstrate that the
first order transition characterised by a jump in the layer
spacing and the two-phase coexistence region terminates at a CP
in the temperature- composition plane. Further, we observe that
the values of $\beta$ and $\delta$, the critical exponents
associated with the order parameter and the course of the
critical isochore are in close agreement with the experimental
results obtained in other layered smectics and also with the DP
predictions for the new universality class. Both the pure
compounds exhibit C and I phases. (Although the compound C10 is
chiral, for the purposes of this paper, we shall ignore the
differences between chiral and non chiral materials). The
partial phase diagram for this binary system is shown in Fig.1.
This diagram has been obtained by observing the optical textures
exhibited by the different phases under a polarising microscope
and also simultaneously monitoring the transmitted intensity and
temperature on a computer. The latter method was particularly
useful for identifying the C--I transition. The twist grain
boundary (TGB) phase was identified by its characteristic vermis
texture\cite{ci9d}.  The C and I phases show the standard bubble
texture. High precision layer spacing measurements were
performed using an aligned sample obtained by cooling at a slow
rate ($\sim$ 1$^o$C/h) from the cholesteric phase in the
presence of a magnetic field (2.4T).  Once the sample was in the
C phase, the temperature controlled oven was shifted to the Xray
goniometer head. The Xray setup used has been described in
detail elsewhere\cite{ci10}. Briefly recalling the essential
details, a Ge monochromator positioned to allow only
K$_{\alpha1}$ radiation and a linear position sensitive detector
using a quartz wire enable the precision in the wave vector
determination to be 2 $\times$ 10$^{-4}~\AA^{-1}$, while the
resolution in the equatorial direction (also the scanning
direction) is 1 $\times$ 10$^{-3}~\AA^{-1}$ HWHM.  The precision
in the determination of the temperature is reckoned to be $\pm$
2mK and during any scan the temperature of the sample was
maintained to a constancy of $\pm$ 10 mK.

Fig.2a shows the layer spacing determined as a function of temperature for
TB10A.  The transformation from C to I is signalled by the
coexistence of the peaks due to both the phases (see insets of Fig. 2a),
exactly as is expected and seen earlier for first order transitions
between two smectic phases with different layer periodicities
\cite{ci11,ci7,ci10}.  In contrast the d variation is smooth for C10
(Fig.2b), which together with the fact that the two phases have the same
symmetry, indicates a continuous evolution. Now, it is quite natural
to expect a CP to exist for a suitable composition of the two compounds.
To locate the CP, Xray studies have been performed on seven
different mixtures, X=0.40, 0.34, 0.32, 0.305, 0.29, 0.27 and
0.25 (where X indicates the weight fraction of TB10A in the
mixture). The thermal variation of d for these mixtures is shown
in Fig.3. It is observed that for X$ >$ 0.27 the transition is
first order accompanied by a jump in $d$, while for X = 0.25 and
beyond there is only a continuous evolution; the composition X =
0.27 shows a vertical inflection. Both $\Delta d$, the jump in
the layer spacing and the width of the  two-phase region are
observed to decrease smoothly to zero as the concentration X =
0.27 is approached (see Fig.4).  All these features show
unambiguously that there is a critical point at or in the
immediate vicinity of X$_c$=0.27.

Since $\Delta d$ serves as the order parameter for the
transition, on approaching CP it is expected to follow
\begin{equation}
\Delta d \propto \mid t \mid ^{\beta}
\end{equation}
with t = (T$_{eff}$ -- T$_c$).
T$_c$ is the critical temperature and T$_{eff}$ is the transition
temperature of the mixture being taken as the mean of the high
and low-temperature limits of the coexistence region.  A plot of
$\Delta d$ as a function of $t$ is also shown in Fig.4  along with
a least-squares fit to eq.(1). The fit yields $\beta$ = 0.51 $\pm$ 0.01. For
smectic layered systems, the path of approach to CP along $X =
X_c$ is analogous to the approach along the critical pressure
for the liquid--gas systems\cite{ci9}. In such a situation for X=X$_c$, the
temperature variation of the layer spacing can be written as.
\begin{equation}
d = d_c + A^ \pm \mid t \mid ^{1/\delta}  +  B(T - T_c)
\end{equation}
Here $d_c$ is the $d$ value at $T_c$, A$^ \pm $ denote the amplitudes
above and below $T_c$ and the linear term describes a background
variation.  A fit to equation (2) for the $d$ variation at $X_c$ = 0.27 is
shown in Fig. 5 and gives a $\delta$ of 1.82 $\pm$ 0.2. The
ratio $\frac{A^{-}}{A^{+}}$ comes out to be --1.06.   Although
the value of $\beta$ (= 0.51 $\pm 0.01$) agrees with the MF
value of 0.5, the $\delta$ value (1.82 $\pm 0.2)$ is appreciably
different from the MF value of 3.  It may be recalled here that
similar features were seen for the $C^*$~--~C critical
point\cite{ci8}. Also, the $\delta$ value agrees closely, within
the experimental errors, with that obtained for the A$_d~--~A_2$
critical point\cite{ci9}. The $\delta$ value is also comparable
to the theoretical value of 2 $\pm$ 0.1 given by DP. However,
the theory suggests that the order parameter variation is
described by the ratio $\beta /\Delta$ rather than $\beta$.
Since $\beta /\Delta$=1/$\delta ~ \sim $0.5, the variation of
$\Delta$d and the temperature dependence of d at X=X$_c$ yield
the same exponent. But it should be noted that the  value $\beta
/\Delta$ for the new universality class is not different from
the meanfield $\beta$ value. In view of this, we suggest that
our results support the concept of universal behaviour of
critical points in layered systems.  Specific heat measurements
on the C--I critical point would be interesting and test the
theoretical predictions further.

Thanks are due to Mr. K.Subramanya for technical assistance. We
are grateful Prof. J.Prost for his useful comments.

\newpage
\centerline {\bf Figure Legends}
\ni
Figure 1: Partial temperature-composition (T-X) phase diagram of
the binary system TB10A and C10. Here X indicates the weight
fraction of TB10A in the mixture. The different phases seen are
cholesteric (Ch), twist grain boundary phase (TGB), smectic C
(C) and smectic I (I). The circles indicate the concentrations
for which Xray studies have been done. Lines are meant to be a
guide to the eye.

\ni
Figure 2(a): Temperature dependence of the layer spacing d for
TB10A showing a two-phase coexistence in the transition region.
The insets show raw Xray line profiles in the i) I phase,
ii) coexistence region and iii) C phase. T$_{eff}$ is the
transition temperature.\\ Figure 2(b): Layer spacing variation
in C10 showing a continuous evolution from C to I phase.
T$_{eff}$ refers to the temperature of the inflection point.

\ni
Figure 3: d versus T plots for different concentrations near the
CP. The concentrations are indicated above each plot. The dashed
lines serve as a guide to the eye through the two-phase region.
The solid line is a locus of the boundary of the two-phase
region for different mixtures. The values of T$_{eff}$ for
different concentrations are X=0.4, 105.94$^o$C; X=0.34,
103.50$^o$C; X=0.32, 102.70$^o$C; X=0.305, 101.36$^o$C; X=0.29,
100.76$^o$C; X=0.27, 99.35$^o$C; X=0.25, 98.62$^o$C.  X=0.27
shows a vertical inflection due to the presence of the critical
point (indicated by a solid circle).
\newpage
\ni
Figure 4: Variation of the jump in the order parameter $\Delta$d
(solid circles) and the width of the coexistence region
(triangles) as the CP concentration X$_c$=0.27 is approached.
The lines are meant to be a guide to the eye. Open circles
indicate the order parameter jump obtained for different
concentrations as a function of T$_{eff}$--T$_c$. Notice that as
T$_{eff}$ approaches T$_c$, the jump decreases to zero.  The
solid line represents the "critical isochore" and is a fit to
eq. (1) yielding $\beta$=0.51$\pm$0.01. The scales for the
different data sets are indicated by arrows.

\ni
Figure 5: Enlarged view of  d versus temperature in the
critical region for X$_c$=0.27. The solid line is a fit to eq.(2)
giving $\delta$=1.82$\pm$0.2.
\end{document}